\documentstyle[preprint,tighten,floats,prd,aps]{revtex}
\begin{document}
\pagestyle{empty}
\draft
\title{
 \centerline{\rm\small July 14, 1994 \hfill hep-ph/yymmxxx}
 \centerline{\rm\small       \hfill FERMILAB--PUB--94/196--T}
 \centerline{\rm\small       \hfill UICHEP--TH/94--9}
 \centerline{\rm\small       \hfill NHCU--HEP--94--17}
\bigskip
{$CP$ violation in the cubic coupling of neutral gauge bosons\bigskip}
      }
\author{
Darwin Chang$^{a,b}$,
Wai--Yee Keung$^{c,d}$ and
Palash B. Pal$^{e}$\\
\quad\\  \quad \\
}
\address{
$^a$ Physics Department,
National Tsing-Hua University,
Hsinchu, 30043 TAIWAN\\
$^b$ Institute of Physics,
Academica Sinica, Nankang,
Taipei, TAIWAN\\
$^c$ Physics Department,
University of Illinois at Chicago,
Illinois 60607--7059 \\
$^d$ Fermilab, P.O. Box 500, Batavia, Illinois 60510\\
$^e$ Indian Institute of Astrophysics, Bangalore 560034, INDIA\\
\quad\\
}
\date{July 14, 1994}
\maketitle
\vskip 1cm

\begin{abstract} \normalsize\noindent
We investigate the $CP$ violating form factor of the $ZZZ$
and $ZZ\gamma$ vertices in the
pair production of $Z^0$ bosons. Useful observables in azimuthal
distributions are constructed to probe $CP$ nonconservation which
may originate from these vertices.
A simple Two Higgs Model of $CP$ violation is used as an illustration.
\end{abstract}
\vfill
%
%
\narrowtext
\pagestyle{plain}
In the near future, with the availability of experimental data at energy
around the electroweak breaking scale, one expects to
learn about the structure of the cubic and quartic self-interactions
of gauge bosons. So far, these interactions have not been directly tested
in any experiments.

One exciting possibility is that such interactions will give new
insights on $CP$ violation, whose physical origin has not been
understood with satisfaction yet.
The observation of $CP$ violation in the kaon system can be
explained in various ways within the framework of gauge theories, and
choosing between them requires additional observation of $CP$ violation.
With this in mind, it is interesting to look for $CP$ violating signals
which may be induced by the self-interactions of gauge bosons.
We discuss here one such
possibility, where the coupling of three neutral gauge bosons has a $CP$
violating term in it.
We first did a model-independent discussion based on the most general form
factors.  Then a simple model, the two Higgs doublet model, is used as an
illustration of how the form factors may arise in a realistic $CP$
violating theory.

\section{Helicity Amplitudes.}
Such $CP$-odd term is indeed allowed in general on fundamental
grounds, as is obvious from the general parametrization of the cubic
coupling of gauge bosons \cite{KiTs73,GG79,hagiwara}. Most theoretical
studies along this direction have been done\cite{r:gounaris,r:Gavela,r:ckp:ww}
only for the process $e^-e^+\rightarrow W^-W^+$.
The effect of $CP$ violation in $e^-e^+\rightarrow Z^0Z^0$
has not been thoroughly carried through\cite{r:Zralek}
and there is a need of detailed analysis.
This motivates us to perform a careful model-independent study.
In Section IV, a simple Two Higgs Model is used as an illustration.
We follow the helicity formalism for the $Z^0$ pair production,
$e^-(\sigma)e^+(\bar \sigma) \rightarrow Z^0 (\lambda) Z^0(\lambda')$,
outlined in Appendix D of Ref.\cite{hagiwara}.
Here we include explicitly effects from the form factors
$f_4$ and $f_5$ which describe the vertex
$V(P)\rightarrow Z(q)Z(q')$
for out-going on-shell $Z^0$ bosons, where the incoming particle $V$
is either another $Z$-boson or a photon:
\begin{equation}
ie \Gamma_{V\rightarrow ZZ}^{\mu\alpha\beta}
=ie\ {s-m_V^2\over M_Z^2}
\left[
      if_4^V(P^\alpha g^{\mu\beta} + P^\beta g^{\mu\alpha})
     +if_5^V\epsilon^{\mu\alpha\beta\rho}(q-q')_\rho
\right] \quad (V=Z,\gamma) \,,
\label{eq:f4f5}
\end{equation}
where $s=P^2$.
Note that $f_4$ term is $CP$-odd. The
$f_5$ term, although $CP$-even,
is included for completeness.
The helicity amplitudes are given by
\begin{equation}
   {\cal M}_{\sigma,\bar \sigma;\lambda,\lambda'} (\Theta)=
4 \sqrt{2} \: e^2 \:
\: d^{{\rm max}(|\Delta\sigma|,|\Delta\lambda|)}_{\Delta \sigma,\Delta \lambda}
                          (\Theta)
\left[
{
(g_{\Delta\sigma})^2 {A}_{\lambda,\lambda'}(\Theta)
\over
4\beta^2\sin^2\Theta+\gamma^{-4}
}
+\sum_{i=4,5} \gamma^2( g_{\Delta\sigma}f_i^Z-f_i^\gamma)
{A}^{(i)}_{\lambda,\lambda'} \right] \;.
\label{eq:prod}
\end{equation}
The kinematic variables are defined as usual,
$\gamma^{-2}=1-\beta^2=4M_Z^2/s$.
The amplitude for the initial helicity configuration
$\bar \sigma =\sigma$ is highly suppressed
due to helicity argument  in the high energy limit $\sqrt{s}\gg m_e$.
Therefore we are only interested in the cases for which
$\Delta \sigma \equiv {1\over 2}(\sigma -\bar\sigma) =\pm 1$.
The relevant Wigner $d$ functions appearing
in Eq.(\ref{eq:prod}) are listed below:
		\begin{eqnarray}
d^2_{1,\pm2}(\Theta)&=&-d^2_{-1,\mp2}(\Theta)=
\pm{1\over2}(1\pm\cos\Theta) \sin\Theta          ,        \nonumber\\
d^1_{1,\pm1}(\Theta)&=& d^1_{-1,\mp1}(\Theta)=
   {1\over2}(1\pm\cos\Theta)                     ,                 \\
d^1_{1,0}(\Theta)&=&-d^1_{-1,0}(\Theta)
=-{1 \over \sqrt{2}}\sin\Theta   .
\nonumber
		\end{eqnarray}
In the standard electroweak
model at the tree level,
the elements ${A}_{\lambda,\lambda'}(\Theta)$ come from the
$t$-channel exchange diagram.  The electron couplings
$g_{\Delta\sigma}$ to the $Z^0$ boson
are specified by
		\begin{eqnarray}
g_-&=&g_L=\left({1\over \sin\theta_W\cos\theta_W}\right)
(-{1\over 2}+\sin^2\theta_W) \,, \nonumber\\
g_+&=&g_R=\left({1\over \sin\theta_W\cos\theta_W}\right)
(\sin^2\theta_W) \,.
		\end{eqnarray}
After simplification, we summarize the result
for various cases $\Delta\lambda=\lambda-\lambda'$ as follows,
		\begin{equation}
\begin{tabular}{||c|c|c|c|c||}
\hline
$\Delta\lambda$ & $\lambda\ \lambda'$
& ${A}_{\lambda\lambda'}(\Theta)$
& ${A}^{(4)}_{\lambda\lambda'}$
& ${A}^{(5)}_{\lambda\lambda'}$  \\  \hline
$\pm 2$ & $\pm\ \mp$ & $-\sqrt{2}(1+\beta^2)$ & 0 & 0  \\
$\pm 1$ & $\pm\ 0  $ & $(1/\gamma)
[\Delta\sigma\Delta\lambda(1+\beta^2)-2\cos\Theta]$ & $+i\gamma\beta$
& $-\Delta\lambda\gamma\beta^2$\\
$\pm 1$ & $0\ \pm$   & $(1/\gamma)
[\Delta\sigma\Delta\lambda(1+\beta^2)-2\cos\Theta]$ & $-i\gamma\beta$
& $-\Delta\lambda\gamma\beta^2$\\
0     & $\pm\ \pm$ & $-(1/\gamma^2)\cos\Theta$ & 0 & 0\\
0     & $0\ 0   $ & $-(2/\gamma^2)\cos\Theta$ & 0 & 0\\
\hline
\end{tabular} \;.
		\end{equation}

When the kinematic variables of the two identical $Z^0$ boson are
interchanged, i.e.,
		\begin{equation}
(\lambda,\lambda') \leftrightarrow (\lambda',\lambda),\quad
\Theta \leftrightarrow \pi-\Theta,\quad
\Phi \leftrightarrow \pi+\Phi \ ,
\label{eq:boson}
		\end{equation}
the amplitude is unchanged because of the Bose symmetry,
if one includes a negative sign coming from
the azimuthal $\Phi$ rotation $\exp(i\Delta\sigma\pi)$.

The usual $CP$ transformation is
		\begin{equation}
(\lambda,\lambda') \rightarrow (-\lambda,-\lambda'),\quad
\Theta \rightarrow \pi-\Theta,\quad
\Phi   \rightarrow \pi+\Phi \ .
\label{e:CPtrans}
		\end{equation}
However, we can simplify this $CP$  transformation by incorporating
the Bose symmetry in Eq.(\ref{eq:boson}).
The resulting $CP$ transformation becomes
		\begin{equation}
(\lambda,\lambda') \rightarrow (-\lambda',-\lambda),\quad
\mbox{$\Theta$, $\Phi$ unchanged.}
		\end{equation}
The situation now becomes very
similar to our previous analysis\cite{r:ckp:ww}
in the process $e^-e^+\rightarrow W^-W^+$.

If $CP$ is conserved (when $f_4$'s are turned off), we have the
following relation for the amplitudes in our phase convention:
\begin{equation}
{\cal M}_{\sigma,\bar\sigma;\lambda,\lambda'} (\Theta)
={\cal M}_{\sigma,\bar\sigma;-\lambda',-\lambda} (\Theta) \;.
\end{equation}
This equality will be destroyed by the presence of $CP$ violating form
factors $f_4$ in channels $(\lambda,\lambda')=(0,\pm)$ or $(\pm,0)$.

\section{Spin-Density Matrices.}

To avoid studying  the complicated event topology in the 4-fermion final
configuration from the decays of the $Z^0$  pair, we concentrate our
attention to the decay of a single $Z^0$. This strategy is equivalent to
the study of the density matrix for one of the $Z^0$ bosons.

We only look at the $Z^0$ boson at the scattering angle
$\Theta$ and temporarily ignore the recoiling one, which is considered
as produced at the scattering angle $\pi-\Theta$.
The polar angle $\psi$ and the azimuthal angle $\phi$ are defined
in the $Z^0$ rest frame for the lepton
$\ell^-$ in the decay $Z^0\rightarrow \ell^-\ell^+$.
We define the axes of the rest frame of $Z^0$ as follows.
The $z$-axis is along the direction of motion of $Z^0$
in the $e^-e^+$ c.m. frame.
The $x$-axis lies on the reaction plane and toward the direction
where $\Theta$ increases.
The $y$-axis is given by the right-hand rule.

The angular distribution of $\ell^-$ from the $Z^0\rightarrow \ell^-\ell^+$
decay is specified by the
the spin density matrix $\rho_{i,j}$ of the $Z^0$ boson.
\begin{equation}
   \rho(\Theta)_{i,j}
      ={\cal N}(\Theta)^{-1}\sum_{\sigma,\bar\sigma,\lambda'}
  {\cal M  }_{\sigma,\bar\sigma;i ,\lambda'}(\Theta)
  {\cal M}^*_{\sigma,\bar\sigma;j ,\lambda'}(\Theta)
\;.
\end{equation}
Here ${\cal N}$ is the normalization such that $\hbox{Tr}\rho=1$.
$\rho$ is hermitian by definition. The normalized distribution
for $\ell^-$ is given by
\begin{eqnarray}
   {dN(\ell^-,\Theta)\over d\phi\, d\!\cos\psi}={1\over 4\pi}{3\over4}
        \sum_{h=\pm} w_h
             \Bigl[
                (1+h \cos\psi)^2\rho(\Theta)_{++}
              + (1-h \cos\psi)^2\rho(\Theta)_{--}
                +2\rho(\Theta)_{00}  \sin^2\psi \nonumber\\
    -2\sqrt{2} \hbox{Re }\rho(\Theta)_{+,0}(1+h\cos\psi)\sin\psi\cos\phi
    +2\sqrt{2} \mbox{Im}\,\rho(\Theta)_{+,0}(1+h\cos\psi)\sin\psi\sin\phi
                                              \nonumber\\
    -2\sqrt{2} \hbox{Re }\rho(\Theta)_{-,0}(1-h\cos\psi)\sin\psi\cos\phi
    -2\sqrt{2} \mbox{Im}\,\rho(\Theta)_{-,0}(1-h\cos\psi)\sin\psi\sin\phi
                                              \nonumber\\
    +2 \hbox{Re }\rho(\Theta)_{+,-}(1-\cos^2\psi)\cos2\phi
    -2 \mbox{Im}\,\rho(\Theta)_{+,-}(1-\cos^2\psi)\sin2\phi \Bigr]\;.
                                               \nonumber
\end{eqnarray}
\begin{equation}
\label{eq:phi}
\end{equation}
The two contributions come from helicity configurations
$\ell^-_R (h=1)$ and $\ell^-_L (h=-1)$, with different weights,
\begin{equation}
w_-=g^2_L/(g_L^2+g_R^2)\;,\quad w_+=g^2_R/(g_L^2+g_R^2)\;,\quad
w_-+w_+=1\;.
\end{equation}
In our present phase convention,
if $CP$ were conserved ({\it i.e.} when $f_4=0$), we would have
the following identities.
\begin{equation}
\rho(\Theta)_{\lambda,\lambda'}=\rho(\pi-\Theta)_{-\lambda,-\lambda'}
\; ,
\label{eq:Gou}
\end{equation}
based on the transformation in Eq.(\ref{e:CPtrans}).
Similar expressions were first noticed in
Ref.~\cite{r:gounaris} on the process $e^-e^+\rightarrow W^-W^+$
and in
Ref.~\cite{r:ckp:tt} on the process $e^-e^+\rightarrow t\bar t$.

\section{$CP$ violating observables}

Under $CP$ conjugation, we change variables
$\Theta \rightarrow \pi-\Theta$,
$\psi \rightarrow \pi-\psi$, and $\phi \rightarrow -\phi$.
The distribution in Eq.(\ref{eq:phi})
is transformed into itself if
we assume $CP$ conservation as in Eq.(\ref{eq:Gou}).
In the presence of the $CP$-violating term $f_4$, our analysis
of $CP$-violating obsevables in Ref.~\cite{r:ckp:ww} can be
easily applied here.

However, as the coupling of $\ell\bar\ell Z^0$
is almost purely axial-vectorial,
there is approximate charge symmetry $C$, which assigns this vertex
even $C$-parity, with the $f_4$ term also even as well.
Any $C$-odd observable will be suppressed.

We find out that the most prominent effect of $CP$ nonconservation
resides in the elements
$(+,-)$ or $(-,+)$ of the spin-density matrix,
		\begin{equation}
\mbox{Im}\,\rho(\Theta)_{+,-} -
\mbox{Im}\,\rho(\pi-\Theta)_{-,+}
=
{32e^4\over {\cal N}(\Theta)}
\sum_{\Delta\sigma=\pm}
(g_{\Delta\sigma})^2
                   \left(\Delta\sigma\right)
\gamma^2(\beta+\beta^3)
\sin^2\Theta
{
\mbox{Re}\,(f_4^\gamma-g_{\Delta\sigma}f_4^Z)
\over
4\beta^2\sin^2\Theta+\gamma^{-4}} \;.
		\end{equation}
This particular location in the density matrix produces the
azimuthal dependence in the form of $\sin2\phi$.
If we integrate $\psi$ and $\phi$
over quadrants, we expect that $CP$ nonconservation
appears in the folded asymmetry, ${\cal A''}(\Theta)$, which is
		\begin{equation}
{
 [dN(\ell,\Theta,\mbox{I+III})+dN(\ell,\pi-\Theta,\mbox{I+III})]
-[dN(\ell,\Theta,\mbox{II+IV})+dN(\ell,\pi-\Theta,\mbox{II+IV})]
\over
 [dN(\ell,      \Theta,\mbox{I+II+III+IV})
+ dN(\ell,\pi-\Theta,\mbox{I+II+III+IV})]
}\;.
\label{eq:Appud}
		\end{equation}
Here the range of the azimuthal angle has been divided into four usual
quadrants I,II,III and IV.
It turns out that this observable ${\cal A''}$
is $C$-even and thus it is not subjected to the suppression from the
approximate $C$ symmetry.

We can show that
		\begin{equation}
{\cal A''}(\Theta) =-{1\over\pi}
   \Bigl(\mbox{Im}\, \rho(\Theta)_{+,-}
    -\mbox{Im}\, \rho(\pi-\Theta)_{-,+} \Bigr)
\;.
\label{eq:Apd}
		\end{equation}
In Fig.~1, we show the $CP$-odd asymmetry in the density matrix versus
the scattering angle $\Theta$ per unit of small Re $f_4^Z$ at various
energies, $\sqrt{s}=200$, 250, and 300 GeV. Observation of this asymmetry is
a genuine signal $CP$ violation, as it is not faked by the final
state interaction.

It is interesting to note that we do not need to know the charge
of $\ell$ as the events are collected over quadrants I+III or II+IV. We
can use this fact to apply our formula even to the larger sample of jet
events from the $Z^0$ pair without tagging the charges of the primary
partons. Our formalism can be easily translated for the process $q\bar
q\rightarrow Z^0Z^0$ in the hadron collider.

\section{Two Higgs Model}
Cubic couplings among neutral gauge bosons do not appear at the tree
level in the standard
model gauge group of SU(2)$_L\times$U(1)$_Y$. But they can be induced
at the loop level.  In the minimal standard model with just one Higgs
doublet, such amplitudes do not have any $CP$
violation even at the one-loop level, as will be clear from our
analysis below.  We therefore
perform the calculation of $CP$ violating effects in these trilinear
couplings when there are two Higgs doublets \cite{MePo91}
present in the model,
which is a popular model in its own right. Among the possibilities
which open up with the two doublets are: spontaneous $CP$ violation
\cite{Lee}, incorporation of the Peccei-Quinn symmetry \cite{PeQu77}
to solve the strong $CP$ problem, and incorporation of supersymmetry.

At the one-loop level, cubic coupling obviously comes from triangle
diagrams. If the internal lines are fermions, no $CP$ violating effect
is generated at the one-loop level, because the $Z$ or photon
couplings with fermions are flavor diagonal and $CP$ conserving. There
are also triangle diagrams with internal $W$ lines. In the
Feynman--t'Hooft gauge, it can be shown that they do not contribute to
the form factors as shown in Eq.\ (\ref{eq:f4f5}). Thus, for our purpose, we
need to calculate only the diagrams involving Higgs bosons in the
loop. Obviously, such diagrams can never involve the antisymmetric
$\varepsilon$-symbol, so one can only obtain a non-zero $f_4^Z$. This
term has been shown to be non-zero for $WWZ$ coupling at the one-loop
level for the model at hand \cite{HMM93}. We want to extend their
calculation for the case of $V^*ZZ$ couplings, where $V^*$ can be
either an off-shell $Z$-boson or photon, and the other two $Z$-bosons
are assumed to be on-shell.

To set up the notation, we call the two Higgs multiplets to be
$\varphi_1$ and $\varphi_2$. Usually, they are assumed to have special
transformation properties with respect to some discrete symmetries in
order to avoid flavor changing neutral currents at the tree level. We
assume that such discrete symmetries are not imposed on the soft terms
in the Higgs potential, otherwise $CP$ violation would be eliminated in
the Higgs sector of the model. Without any loss of generality, we can
take the vacuum expectation values (VEVs) of $\varphi_1$ and
$\varphi_2$ to be $v_1 \exp(i\vartheta)$ and $v_2$. One can then
define a linear combination  $\varphi$ of the two multiplets which has
a VEV $v=\sqrt{v_1^2+v_2^2}$, and the orthogonal one, $\varphi'$, has
a vanishing VEV. The components of these doublets can then be written as
	\begin{eqnarray}
\varphi = \left( \begin{array}{c} w^+ \\ {1\over \sqrt 2} (v + \phi_1 + iz)
\end{array} \right) \,,   \qquad
\varphi' = \left( \begin{array}{c} H^+ \\ {1\over \sqrt 2} (\phi_2 +
i\phi_3) \end{array} \right) \,.
	\end{eqnarray}
The fields shown here are complex combinations of the fields in the
$\varphi_1$-$\varphi_2$ basis.
The components $w^\pm$ and $z$ are eaten up by the gauge bosons and
disappear from the physical spectrum. There are four physical spinless
bosons in the model. One of them is the complex field $H^+$. The other
three are, in general, superpositions of the fields $\phi_1$,
$\phi_2$, and $\phi_3$. We define the eigenstates by $H_A$, where
	\begin{eqnarray}
\phi_a = \sum_{A=1}^3 O_{aA} H_A \,,
\label{phi=Oh}
	\end{eqnarray}
$O$ being an orthogonal mixing matrix.

The coupling of these neutral Higgs bosons with the $Z$-boson looks
very simple in the $\phi$-basis:
	\begin{eqnarray}
\begin{tabular}{c|l}
Vertex & \multicolumn{1}{|c}{Feynman rule} \\
\hline
$\phi_1 (p) \stackrel{Z_\mu}{\longrightarrow} z(q)$
& ${g\over 2\cos \theta_W} (p+q)_\mu$ \\
$\phi_2 (p)  \stackrel{Z_\mu}{\longrightarrow} \phi_3(q)$
& ${g\over 2\cos \theta_W} (p+q)_\mu$ \\
\end{tabular}
	\end{eqnarray}
Using Eq.\ (\ref{phi=Oh}), it is trivial to rewrite these Feynman
rules in terms of the mass eigenstates of neutral Higgs bosons:
	\begin{eqnarray}
\begin{tabular}{c|l}
Vertex & \multicolumn{1}{c}{Feynman rule} \\
\hline
$H_A (p)  \stackrel{Z_\mu}{\longrightarrow} z(q)$ &
${g\over 2\cos \theta_W} O_{1A}
(p+q)_\mu$ \\
$H_A (p)  \stackrel{Z_\mu}{\longrightarrow} H_B (q)$
& ${g\over 2\cos \theta_W}
(O_{2A} O_{3B} - O_{2B} O_{3A}) (p+q)_\mu$ \\
\end{tabular}
\label{HHZ}
	\end{eqnarray}
Using the orthogonality of the mixing matrix $O$, we can write
	\begin{eqnarray}
O_{2A} O_{3B} - O_{2B} O_{3A} = \sum_C \epsilon_{ABC} O_{1C} \,,
\label{OXO}
	\end{eqnarray}
which simplifies the form of the $Z$-coupling with two physical Higgs
bosons. Notice that the $Z$-coupling between two physical Higgs bosons
is necessarily flavor-changing, which opens up the possibility for $CP$
violation at one-loop level. For the reason that the photon field
preserves flavors at the tree level, there is no
$f_4^\gamma$ form factor at the one-loop calculation
in the Two Higgs Model.

These cubic couplings appear in the triangle diagrams shown in
Fig.~2.
Notice that, in the figure, the Higgs boson
lines have been denoted with subscripts $i,j,k$, which run from 0 to
3, where $H_0$ is identified with the unphysical Higgs $z$ which
appears as intermediate lines since we adopt the Feynman-t'Hooft
gauge. A straightforward calculation now shows that the form-factor
$f_4^Z$ from these diagrams can be written in the form
\begin{eqnarray}
ef_4^Z = - {1 \over 128 \pi^2}
                     \left({e\over\sin\theta_W\cos\theta_W}\right)^3
                            {M_Z^2\over P^2-M_Z^2}
                   \sum_{i,j,k} \lambda_{ijk} I(M_i,M_j,M_k) \,.
\label{eq:f4}
	\end{eqnarray}
Here, $\lambda_{ijk}$ is a factor coming from vertices which will be
discussed below, and the loop integral $I(M_i,M_j,M_k)$ is equal to:
	\begin{eqnarray}
 2!\int\int
(x-y) \; \ln {\Lambda^2 \over xM_i^2 + yM_j^2 + wM_k^2 -
w(1-w)M_Z^2 - xyP^2-i0^+}  dx dy \,,
\label{I}
	\end{eqnarray}
where the positive Feynman parameters $x$ and $y$ are restricted within
the integration domain $x+y \le 1$ and also $w=1-x-y$.
$\Lambda$ is a cut-off which disappears in the expression for $f_4^Z$, as
we will show below. When one of the particles denoted by $i$, $j$ or $k$
is the unphysical Higgs boson, the corresponding mass should be
interpreted to be $M_Z$, because the propagator of the unphysical Higgs
boson has a pole for this value of mass in the gauge we use. For future
purposes, notice that
	\begin{eqnarray}
I(M_i,M_j,M_k) = - \, I(M_j,M_i,M_k) \,,
\label{I-I}
	\end{eqnarray}
which follows from the definition in Eq.~(\ref{I}).

Let us now discuss the factor $\lambda_{ijk}$. First, consider the
case when all the Higgs bosons in the loop are physical ones.
Due to the antisymmetry of the coupling of $H_AH_BZ_\mu$ from
Eq.\ (\ref{HHZ}), all the Higgs bosons in the
loop must be different. If, following the direction of the momentum
arrow in
Fig.~2, we encounter the mass eigenstates
$H_1$, $H_2$ and $H_3$ in that order, it is easy to see that the
factor coming from the vertices is
	\begin{eqnarray}
   \lambda_{123} = O_{11} O_{12} O_{13}\equiv \lambda  \,.
	\end{eqnarray}
Obviously, there are three such diagrams, and their total
contribution is
	\begin{eqnarray}
\lambda \left\{ I(M_1,M_2,M_3) + I(M_2,M_3,M_1) + I(M_3,M_1,M_2)
\right\} \,.
\label{ABCcontrib}
	\end{eqnarray}
On the other hand, if we encounter the eigenstates in the reverse
order, we obtain a factor $-\lambda$ from the vertices.
However, this term will be multiplied by
$$\{ I(M_2,M_1,M_3) + I(M_3,M_2,M_1) + I(M_1,M_3,M_2) \}\ . $$
By virtue of Eq.\ (\ref{I-I}), the product of the two is
the same as the contribution of Eq.\ (\ref{ABCcontrib}).

Next we consider diagrams where one of the internal lines is the
unphysical neutral Higgs boson $z$. Note that since there is no
coupling of the $Z$-boson with two unphysical Higgs bosons, at most
one internal line can be the unphysical Higgs boson. In this case,
one can derive that
	\begin{eqnarray}
   \lambda_{120} =\lambda_{230} =\lambda_{310} =-\lambda \,,
	\end{eqnarray}
and the same value for any even permutation of subscripts, but
opposite sign for an odd permutation. Therefore,
the last factor of summation in
Eq.\ (\ref{eq:f4}) becomes
         \begin{eqnarray}
\sum_{i,j,k} \lambda_{ijk} I(M_i,M_j,M_k) =
2 \lambda \Big\{&+I(M_1,M_2,M_3) + I(M_2,M_3,M_1) + I(M_3,M_1,M_2)& \cr
              &-I(M_1,M_2,M_Z) - I(M_2,M_3,M_Z) - I(M_3,M_1,M_Z)& \cr
              &-I(M_Z,M_1,M_2) - I(M_Z,M_2,M_3) - I(M_Z,M_3,M_1)& \cr
              &+I(M_Z,M_1,M_3) + I(M_Z,M_2,M_1) + I(M_Z,M_3,M_2)&
        \Big\} \,.
\label{eq:Ilist}
         \end{eqnarray}
One can see that the cutoff $\Lambda$ dependence is cancelled by pairs in
Eq.\ (\ref{eq:Ilist}).
We also note that $f_4^Z$ remains finite when $P^2=M_Z^2$ as noted in
Ref.\cite{hagiwara}.

Fig.~3 shows the extremely tiny size ($\sim 10^{-6}$)  of $f_4^Z$ for
typical choices of parameters.
We only use this Two Higgs Model as an illustration
how $CP$ violation occurs even in a purely bosonic sector.

\section{Conclusion}
At LEP II, the $Z^0Z^0$ production cross-section is about
1 pb (See Fig.~4) for $\sqrt{s} = 200$ GeV  which can provide about
$500$ $Z^0Z^0$ pairs per year for the design luminosity of
$5 \cdot 10^{31}$ cm$^{-2}$ s$^{-1}$.
As we have shown in the paper, it is possible to test $CP$ symmetry in
purely charged leptonic, purely hadronic or mixed channels of the
two $Z^0$ boson decays. We may require that at least one of the $Z^0$
decays into the charged leptons in order to avoid backgrounds from the
$W^+W^-$ production. The branching ratio of a single $Z^0$ decaying  into
all charged leptonic channels ($e^+e^- + \mu^+\mu^- + \tau^+\tau^-$)
is about 10\%.
While the event statistics probably will not be large enough to test
some of the popular alternative gauge models of $CP$ violation, it is
nevertheless sufficient to provide nontrivial constraints on the $CP$-odd
form factors in the three gauge boson couplings.

The research of WYK was supported in part by the
U.S. Department of Energy.

\vfill
\centerline{FIGURE CAPTIONS}
\begin{itemize}
\item[Fig.~1]
The $CP$-odd asymmetry in the density matrix
versus the scattering angle $\Theta$ per unit
of Re $f_4^Z$ at various energies, $\sqrt{s}=200$, 250, and 300 GeV.
\item[Fig.~2]
Triangle diagrams with internal scalar lines which give
rise to the $Z^*ZZ$ coupling.
\item[Fig.~3]
The size of $f_4^Z/O_{11}O_{12}O_{13}$ versus the lightest
Higgs mass at $\sqrt{s}=200$ GeV, for the case $M_2$=150 GeV, and
$M_3=250$ GeV. The real and the imaginary parts are given by the
solid and the dashed lines respectively.
\item[Fig.~4]
Differential cross-section $d\sigma/d\cos\Theta$ for
$e^-e^+\rightarrow Z^0Z^0$ at various energies $\sqrt{s}$=
200 (solid), 250 (dashed), and 300 GeV (dased-dotted),
predicted by the Standard Model.
The horizontal lines indicate the level of the
corresponding total cross-sections.
\end{itemize}
\end{document}